\documentclass[11pt,reqno]{amsart}

\usepackage{amsmath,amsfonts}
\usepackage{latexsym,amssymb}
\usepackage[dvipsnames]{xcolor}
\usepackage{graphicx}

\newcommand{\dd}{\mathtt{d}}

\begin{document}

\title{On the ambiguity of solutions of the system of the Maxwell equations.}

\author{Vladimir Onoochin}

\maketitle

\begin{abstract}

This work, that is devoted to the memory of Dr. Andrew Chubykalo and his legacy,  is the improved version of the paper published in {\it Annales de la Fondation Louis de Broglie}~\cite{AFLB}. 

In this article, methods for solving the Maxwell equations are analyzed. First, a method based on the direct differentiation of Maxwell's equations and their reduction to the wave equation for an electric field is considered in detail. It is shown that this method cannot give a closed-form solution that does not contain integrals.
 
The other method, which is most often used by scientists, is based on the introduction of auxiliary quantities such as electromagnetic potentials. But rewriting the Maxwell equations via potentials requires the introduction of an additional constraint - the gauge condition. Some analysis of the expressions for the electric field obtained in two gauges, Coulomb and Lorenz gauges, shows that, contrary to the generally accepted opinion that expressions for the electromagnetic field should be identical in any gauge, the final expressions for the electric field differ in different gauges. Consequently, the uniqueness of the solution of the Maxwell equations is absent.  
\end{abstract}

\section{Introduction}

Although the Maxwell equations are the most well-known and most used equations in physics, perhaps due to obvious reliability of these equations, some of their aspects remain outside consideration. One of such aspects is a problem of uniqueness of solutions of the Maxwell equations. 

It should be noted that the proof of this uniqueness is given in some textbooks, for example, in '{\it Electromagnetics}'  of Rothwell and Cloud~\cite{Roth} and '{\it Modern electrodynamics}' of Zangwill~\cite{Zang}. To the author's knowledge, first analysis of uniqueness of the Maxwell equations has been given in the textbook of Tamm of 1929~\cite{Tamm} who assumes that two solutions ${\bf E}_1,\,{\bf H}_1$ and ${\bf E}_2,\,{\bf H}_2$ are derived from the same charge and current distribution. Considering the difference of these solutions, Tamm shows that if this difference is not equal to zero, it will be in contradiction to the Poynting theorem. Other authors follow Tamm's proof. But all authors omit a description how these solutions have been obtained. Meanwhile derivation of the expressions for the EM field from the Maxwell equation is not a trivial procedure. 
 
The physicists write the Maxwell equations to describe electrodynamic systems but these equations are not solved to find the $E$ and $H$ fields. Instead, as O'Rahilly notes (p.~184 of~\cite{O'}), everyone uses the retarded potentials introduced by Lorenz~\cite{LL} to obtain the solution. Lorenz also showed that the wave equations introduced by Riemann~\cite{Rie} and the condition on the potentials, now called the gauge condition, are equivalent to the Maxwell equations.

This approach to solving these equations is entirely reasonable. Despite the fact that the Maxwell equations are linear, two of the equations in the system contain both time and coordinate partial derivatives, while the other two contain only coordinate partial derivatives. Therefore it looks like a natural way to reduce the system of four equation to the equation containing the only unknown, the scalar or vector potential. 

However, no one has shown the opposite - that the Maxwell equations are equivalent to the Riemann-Lorenz wave equations? Meanwhile, if the wave equations give a unique solution, the problem: whether the Maxwell equations have a unique solution, remains out of the question. Despite the apparent absurdity of asking such a question, since there is no reason to doubt the validity of the Maxwell equations, it can be stated.  

First, the question is not about the validity of the Maxwell equations, but about the possible existence of several solutions of this system. Second, when solving this system of equations by means of the potentials, an additional condition is introduced which connects these potentials (the gauge condition). In fact, this condition is introduced arbitrarily, since in the modern interpretation of electrodynamics the potentials are treated as a mathematical tool.  But if the gauge condition is introduced arbitrarily, and therefore the potentials are also determined with a certain degree of arbitrariness, it is reasonable to ask whether any sets of potentials defined by this condition lead to the same expressions for the fields. The latter is necessary because the fields are solutions of Maxwell's equations. And if the expressions for the fields are different for different gauge conditions, it can be concluded that the system of Maxwell equations has multiple solutions for the EM fields. 

One may object that the method of solving the Maxwell equations without introducing potentials is well known, the method developed by Jefimenko~\cite{Jef}. In this case, the wave equations for the fields $E$ and $H$ are derived directly from the Maxwell system. With such a derivation, the introduction of a gauge condition is not required. Consequently, the (possible) problem of differences in the expressions for the EM fields in different gauges is absent. However, the procedure for deriving the wave equation for the field $E$ contains one incorrect mathematical assumption. Let us show this, considering simplest  electrodynamic system, namely, the classical (point) charges moving in a vacuum in an arbitrary way and creating the EM fields. 

\section{Derivation of the wave equation for the electric field}

To solve the system of Maxwell equations, it is necessary to reduce these equations to an equation whose solution is known. The partial differential equations (PDEs) with known solutions are the Poisson equation and the wave equation. However, the Poisson equation contains only spatial variables. Meanwhile, two of the Maxwell equations contain the partial time derivatives. Therefore, it is reasonable to reduce the Maxwell equations to the wave equation, since the latter contains the time derivative. This can be done either by introducing the potentials (a method dating back to Lorenz), or by directly differentiating the initial equations for the EM fields. 

Let us consider the derivation of the wave equation for the electric field without the introduction of the potentials. The starting point is two Maxwell equations (the Gauss units will be used throughout this article; $\varepsilon =\mu=1$ for the vacuum) 
\begin{eqnarray}
&&\frac{\partial {\bf H}}{c\partial t}=-\nabla\times{\bf E}\,; \label{Mxw1}\\
&&\nabla\times{\bf H}=\frac{\partial {\bf E}}{c\partial t}+\frac{4\pi{\bf j}}{c} \,. \label{Mxw2}
\end{eqnarray}
Taking the curl of Eq. (\ref{Mxw1}) and the time derivative of Eq. (\ref{Mxw2}), one has
\begin{eqnarray*}
&&\nabla\times \frac{\partial {\bf H}}{c\partial t}=-\left[\nabla\times[\nabla\times{\bf E}]\right]\,;\\
&&\nabla\times  \frac{\partial {\bf H}}{c\partial t}=\frac{\partial ^2{\bf E}}{c^2\partial t^2}+\frac{4\pi\partial{\bf j}}{c^2\partial t}\,.
\end{eqnarray*}
Since the {\it lhs}'s of these equations are equal, their {\it rhs}'s should be also equal. It gives 
\begin{equation*}
-\left[\nabla\times[\nabla\times{\bf E}]\right]=\frac{\partial ^2{\bf E}}{c^2\partial t^2}+\frac{4\pi\partial{\bf j}}{c^2\partial t}\,,
\end{equation*}
or presenting in a form more acceptable to mathematicians when the source is in the {\it rhs} of the equation, and the differential operators are in its {\it lhs}
\begin{equation}
-\left[\nabla\times[\nabla\times{\bf E}]\right]-\frac{\partial ^2{\bf E}}{c^2\partial t^2}=\frac{4\pi\partial{\bf j}}{c^2\partial t}\,.\label{nwe}
\end{equation}
The above expression is a differential equation for the $E$ field. This equation is typically treated as a differential equation that does not contain the longitudinal component of the radiated E field. This is because the operator $\nabla\times$ "cuts" it from the equation. However, Eq.~(\ref{nwe}) is not yet in the form of a wave equation. 

It should be noted that according to Jefimenko, the solution of this equation is also a retarded integral, Eq.~(2-1.13) of~\cite{Jef}. In the other words, Eq.~(\ref{nwe}) is some modification of the wave equation. But it is not the case, that is shown in {\it Appendix}.

Meanwhile, to reduce Eq.~(\ref{nwe}) to the wave equation (of the true form), it is necessary to use the vector identity 
\begin{equation*}
-\left[\nabla\times[\nabla\times{\bf E}]\right]=\nabla^2{\bf E}-\nabla\left(\nabla\cdot{\bf E}\right)\,,\label{vi}
\end{equation*}
and then remove the last term of that identity. Removal of this term is possible using one more Maxwell equation 
\begin{equation}
\nabla\left(\nabla\cdot{\bf E}\right)=4\pi\nabla \rho  \,.\label{nwm}
\end{equation}
Indeed if one makes term-by-term summation of Eqs. (\ref{nwe}) and (\ref{nwm}), it gives,
\begin{eqnarray}
-\left[\nabla\times[\nabla\times{\bf E}]\right]-\frac{\partial ^2{\bf E}}{c^2\partial t^2}\,\,&=&\,\,\frac{4\pi\partial{\bf j}}{c^2\partial t}\,;\nonumber\\
{\bf +}\quad\quad\quad\nabla\left(\nabla\cdot{\bf E}\right)\,\, &=&\,\, 4\pi \nabla\rho\,;\nonumber \\
\quad \Longrightarrow\quad
\nabla^2{\bf E}-\frac{\partial ^2{\bf E}}{c^2\partial t^2}\,\,&=&\,\,4\pi \left( \nabla \rho+\frac{\partial{\bf j}}{c^2\partial t}\right)\,.\label{fE}
\end{eqnarray}
Thus, Eq.~(\ref{fE})  takes a form of Jefimenko's wave equation for the electric field.

However, there is certain problem with this equation. To obtain Eq.~(\ref{fE}), one should make summation of two PDEs' of different types -- Eq.~(\ref{nwe}) is of the hyperbolic type and Eq.~(\ref{nwm}) is of the elliptic type. Strictly speaking, there is no theorem in the theory of partial differential equations that allows doing such a term by term addition. 

A slightly different approach is used by Richardson (Ch. IX of~\cite{Rich}; Jefimenko mentions the procedure for deriving the wave equation for Richardson's electromagnetic fields). In considering Eq.~(\ref{nwe}), after applying the vector identity, {\it i.e.} the equation
\begin{equation*}
\nabla^2{\bf E}-\nabla\left(\nabla\cdot{\bf E}\right)-\frac{\partial ^2{\bf E}}{c^2\partial t^2}=
\frac{4\pi\partial{\bf j}}{c^2\partial t}\,,\label{nwe2}
\end{equation*}
 this author states that $\nabla\left(\nabla\cdot{\bf E}\right)$ can be changed to $4\pi \nabla\rho$ which should be true because of the Maxwell equation.

There is, however, one objection to this change -- in fact, Richardson replaces the equation with its {\it rhs}. However, this is an incorrect change, since the equation cannot be changed to its source. The source ($\rho$) can create fields, but the fields cannot create the source, the classical electron. 

Thus, the above analysis shows that in the procedure of solution of the Maxwell equations, the wave equation for the fields cannot be derived in the rigorous way. In the other words, solving the Maxwell equations is possible only {\it by introducing the potentials}.
 
 \section{Solving the Maxwell equations by introducing the potentials}
 
The second method to solve the Maxwell equations is to rewrite the EM fields via the potentials ${\bf A}$, $\varphi$, and then to find the solution of the equations for ${\bf A}$ and $\varphi$ .

It should be noted that in modern classical electrodynamics, electromagnetic potentials are considered to be "fictitious" or abstract mathematical quantities that have no physical meaning. However, in the mid-XIX century, when wave equations for EM potentials were derived, these potentials came to be regarded as quantities equivalent to EM fields.
The original set of equations presented by Maxwell in his 1865 paper~\cite{M} and later in his {\it Treatise} are equations not only for electromagnetic fields, but also for potentials, currents, and the forces acting on these currents. However, these equations were presented in the so-called `physical form', that is, as relations, rather than in the classical form used in the theory of partial differential equations (PDEs), where the differential operators are on the {\it lhs} and the sources on the {\it rhs} of the equations. Moreover, Maxwell did not distinguish between `cause and effect'. Obviously, it was practically impossible to obtain solutions to such equations. Maxwell's followers, the so-called {\it Maxwellians}, {\it i.e.} Heaviside, Fitzgerald, Lodge and Hertz significantly reduced the number of original equations. First, they eliminated forces from the original set of Maxwell's equations, and second, they eliminated potentials. After they simplified the system of Maxwell's equations, potentials are used as auxiliary quantities for determining EM fields, 
\begin{equation}
{\bf E}=-\nabla\varphi -\frac{1}{c}\frac{\partial {\bf A}}{\partial t}\,;\quad {\bf H}=\nabla\times{\bf A}\,. \label{P}
\end{equation}
Although Maxwell's original system of 20 equations had essentially been reduced to a system containing only four equations, a general method for solving this system of partial differential equations is still unknown. 

It should be noted that Maxwell derived equations for the components of the vector potential in the form of a homogeneous wave equation (Article 784 of his {\it Treatise}), but without solving these equations. Based on their form, Maxwell concluded that light should propagate as magnetic field waves. It had already been established in the XIX century that the propagation of light is a wave process. Hertz experimentally confirmed that the propagation of EM fields is also a wave process. Therefore, it was quite reasonable that in order to solve the Maxwell equations, they should be reduced to wave equations. 

Actually, the aim of such a reduction is not derivation of the wave equation but presenting the Maxwell equations containing at least two unknowns, ${\bf E}$ and ${\bf H}$ , to the equations which contain the only unknown. It is achieved by rewriting two Maxwell equations via the potentials by means of (\ref{P}),
\begin{equation}
\begin{split}
&[\nabla\times{\bf H}] -\frac{1}{c}\frac{\partial {\bf E}}{\partial t}=\frac{4\pi}{c}{\bf j}\,\,\,\,\to\,\,\,\,
\left[\nabla\times[\nabla\times{\bf A}] \right] +\frac{1}{c^2}\frac{\partial^2 {\bf A}}{\partial t^2}
+\frac{1}{c}\nabla\frac{\partial \varphi}{\partial t} =\frac{4\pi}{c}{\bf j}\,;\\
&\nabla\cdot{\bf E}=4\pi\rho \,\,\,\,\to\,\,\,\,-\frac{1}{c}\frac{\partial }{\partial t}\left(\nabla\cdot{\bf A}\right)
-\nabla^2\varphi = 4\pi\rho\,.
\end{split} \label{eqsP}
\end{equation}
Although both Eqs.~(\ref{eqsP}) have two unknowns, the latter can be easily separated by imposing a condition on the potentials
\begin{equation}	
\frac{1}{c}\frac{\partial \varphi}{\partial t}+\nabla\cdot{\bf A}=0\,,\label{gauge}
\end{equation}	
Then one obtains two wave equations for the potentials,
\begin{equation}
-\nabla^2{\bf A}+\frac{1}{c^2} \frac{\partial^2{\bf A}}{\partial t}=\frac{4\pi{\bf j}}{c}\,\,;\,\,
-\nabla^2 \varphi +\frac{1}{c^2} \frac{\partial^2\varphi}{\partial t}= 4\pi \rho\,. \label{WaEq}
\end{equation}
It should be noted that to derive the wave equations for the potentials, only {\it two} Maxwell's equations plus condition~(\ref{gauge}) are needed but not the whole set of the Maxwell equations. 

This condition, called the Lorenz gauge condition, is a consequence of the charge conservation law. Indeed, if one writes the potentials as Lorenz's retarded integrals of the charges and currents~\cite{LL}
\begin{equation}
\varphi({\bf r},t)=\int\frac{[\rho]_{ret}}{|{\bf r}-{\bf r}'| }\dd {\bf r}'\,;\quad {\bf A}=\frac{1}{c}\int\frac{[{\bf j}]_{ret}}{|{\bf r}-{\bf r}'|}\dd {\bf r}'\,,	\label{Lret}
\end{equation}
one obtains
\begin{equation}	
\frac{1}{c}\frac{\partial \varphi}{\partial t}+\nabla\cdot{\bf A}=\frac{1}{c}\int \frac{[\partial_t\rho]_{ret}}{|{\bf r}-{\bf r}'|}\dd {\bf r}'+\frac{1}{c}\nabla\cdot\int \frac{[{\bf j}]_{ret}}{|{\bf r}-{\bf r}'|}\dd {\bf r}'=\frac{1}{c}
\int \frac{\left[\partial_t \rho+\nabla\cdot{\bf j} \right]_{ret}}	{|{\bf r}-{\bf r}'|}\dd {\bf r}'=0\,,\label{chC}
\end{equation}
where the relation
\[
\nabla\cdot \frac{[{\bf j}]_{ret}}{|{\bf r}-{\bf r}'|}= \frac{[\nabla'\cdot{\bf j}]_{ret}}{|{\bf r}-{\bf r}'|}
\]
is proven in Ch. 1-2 of~\cite{Jef}.

Solutions of Eqs.~(\ref{WaEq}) give the potentials in the Lorenz gauge. The electromagnetic fields are found from these potentials by means of (\ref{P}).

However, the potentials in Eqs.~(\ref{eqsP})  can be separated in another way, namely by using Maxwell's condition $\nabla\cdot{\bf A}=0$ (Art. 783 of Maxwell's {\it Treatise'}), which is now defined as the Coulomb gauge. Then the term containing  $\nabla\cdot{\bf A}=0$ in the second of Eqs.~(\ref{eqsP}) vanishes, and this equation is transformed to the Poisson equation 
\begin{equation*}
-\nabla\left(\nabla\varphi +\frac{1}{c}\frac{\partial {\bf A}}{\partial t}\right)=4\pi\rho\,\, \to\,\,
	-\nabla^2\varphi=4\pi\rho\,.
\end{equation*}
The solution of this equation is known, it is the scalar potential that propagates instantaneously. Having known the solution for the scalar potential, one can treat the term $\nabla\left(\partial\varphi /\partial t\right)$ in the first of Eqs.~(\ref{eqsP}) as a source along with the term containing the current density. In this case it is easy to obtain the following wave equation for the vector potential

\begin{equation}
-\nabla^2{\bf A}+\frac{1}{c^2} \frac{\partial^2{\bf A}}{\partial t}=\frac{4\pi{\bf j}}{c}-\nabla\frac{\partial\varphi}
{c\partial t}\,. \label{weqC}
\end{equation}
Thus, the application of two gauge conditions yields two extreme cases of the types of propagation of scalar potentials. If the vector potential obeys the wave equation in both cases, the scalar potential propagates either with the speed of light (the Lorentz gauge) or with infinite speed (the Coulomb gauge). Moreover, assuming the relation between the potentials, 
\begin{equation}
\nabla \cdot{\bf A}+\frac{c}{u^2}\frac{\partial \varphi }{\partial t}=0\,,\label{gau}
\end{equation}
it is possible to obtain separate (wave-) equations for the scalar potential and the vector potential; the latter, however, contains derivatives of the scalar potential as a source~\cite{McD}. Since the parameter $u$ can vary continuously in the range from $u=c$ to $u=\infty$, one can obtain an infinite number of equations for the potentials. 

The question arises: does the existence of an infinite number of systems of equations for potentials mean the existence of an infinite number of solutions for EM fields, or do all these systems of equations give the same expressions for the EM fields? As shown in Sec. ~2, the Maxwell equations can only be solved by introducing potentials, so the question can be reformulated: do the Maxwell equations have a unique solution for the EM fields, or do they have an infinite number of solutions (corresponding to an infinite number of possible gauge conditions)? 

Let us note that Eq.~(\ref{gau}) contains no physical justification -- it is a purely mathematical constraint, and its introduction became possible after the work of Heaviside, when the potentials began to be treated as mathematical symbols. Meanwhile, both the Lorenz and Coulomb gauge conditions were initially considered as a link between the physical quantities. The Lorenz gauge can be considered as a consequence of the charge conservation law, Eq.~(\ref{chC}). Maxwell uses certain physical arguments to introduce the Coulomb gauge. Strictly speaking, he introduces the condition div${\bf A}=J$ (where $J$ is not the current density) and then states that $J$ will be zero everywhere ( Art. 616 of~\cite{M}). 

Obviously, the problem of uniqueness of solutions of Maxwell's equations can arise if the potentials calculated in different gauges give different expressions for the EM fields. Therefore, a number of papers have been published in which the authors present proofs of the equivalence of these fields; the first paper in this series is the paper by Brill and Goodman~\cite{BrG}. But it is easy to see that all these papers, with the exception of a single one (by Hnizdo~\cite{Hn}), do not contain {\it any example} of explicit expressions for the fields $E$ or $H$ calculated in the two most used (Lorenz and Coulomb)  gauges. 

The absence of these examples could be acquitted by cumbersome computations. But such computations can be done in a case of a single charge which set suddenly from rest into uniform motion. In~\cite{R-G} this example -- with detailed calculations -- is given. This example yields the difference in the electric field which says in favor of the absence of the gauge equivalence. As a consequence, existence of this difference means that a set of the Maxwell equations has two solutions. 

It can be said that it is not enough to have one counterexample to gauge equivalence - since publication of Brill and Goodman's paper, there have been a number of papers proving gauge equivalence. Although no one of these papers contains a direct closed-form calculation of the electric field, it is difficult to think that all of these papers contain errors. It would be inappropriate to analyze their contents in search of weak points. But one common flaw in them can be pointed out. This flaw casts doubt on the presented proof of gauge equivalence. 

\section{Problems with computation of electric fields in different gauges}

In this section it will be described the obstacles to demonstrating that the expressions for the field $E$ are identical in all gauges. Such a demonstration at least requires calculating this field in two gauges. But due to the term $\nabla\dfrac{\partial \varphi_C}{\partial t}$, the solution of Eq.~(\ref{weqC}) is represented by a retarded integral of a nonlocal source. For the gauge (\ref{gau}) a similar retarded integral also appears. 

This retarded integral cannot be computed in the general case. Therefore, the only possibility to prove the equivalence of gauges and, as a consequence, uniqueness of solutions of the Maxwell equations is to transform  the expressions for $\varphi_C,\,{\bf A}_C$ and $\varphi_L,\,{\bf }_L$ to identical forms (when the expressions for $\varphi_L,\,{\bf A}_L$ or their derivatives have the same forms as the expressions for $\varphi_C,\,{\bf A}_C$ or their derivatives). In most works on this subject, the authors do the same.

But the identity of the forms does not guarantee this. In the general case, these forms are represented as integrals whose evaluation takes place in some integration domain. Due to properties of the scalar potential in two of these gauges, the integration domains are different for these gauges. Obviously, even if the integrals have identical forms but are evaluated in different domains, their final values cannot be equal. This is the main obstacle in any proof of the equivalence of the gauges. Let us show this in more detail.  

Assuming the electric fields in the Lorenz and Coulomb gauges are equal, let us rewrite the difference between these fields in terms of potentials 
\begin{eqnarray*}
\nonumber
{\bf E}_{\rm L}-{\bf E}_{\rm C} = 0& \quad
\Longrightarrow \\
-\nabla\varphi_L+\nabla\varphi_C-\frac{1}{c}
\frac{\partial {\bf A}_L}{\partial t}+\frac{1}{c}
\frac{\partial {\bf A}_C}{\partial t}=0&\,\,.\label{eqv}
\end{eqnarray*}
Because the vector potential in two gauges obeys the following equations,
\begin{eqnarray*}
\nabla ^2\mathbf{A}_C-\frac{\partial ^2\mathbf{A}_{C}}{c^2\partial t^2} = -\frac{4\pi }{c}\mathbf{J}
+\frac{1}{c}\nabla \frac{\partial \varphi _{C}}{\partial t}\,,\\
\nabla ^2\mathbf{A}_{L}-\frac{\partial ^2 \mathbf{A}_{L}}{c^{2}\partial t^2}= -\frac{4\pi }{c}\mathbf{J}\,,
\end{eqnarray*}
the difference of the vector potential in two gauges  becomes
\begin{equation}
\mathbf{A}_{\Phi}=\mathbf{A}_C-\mathbf{A}_L=-\int \frac{1}{|{\bf R}-{\bf r}|}
\left[ \nabla_r\frac{\partial \varphi _{C}({\bf r};t)}{c\partial t}\right] \,\dd ^3 r\,, \label{dA}
\end{equation}
where the square brackets [...] are used for the quantities depending on the retarded time $t_{ret}=t- |{\bf R}-{\bf r}|/c$. Thus, the equivalence of the gauges will take place if the relation
\begin{equation}
-\nabla\varphi_L+\nabla\varphi_C+\frac{1}{c}\frac{\partial {\bf A}_\Phi}{\partial t}=0 \,,\label{Rel}
\end{equation}
is true. 

As it is mentioned above, in most works on this subject the fulfillment of this or similar relation is proved. Let us consider how it is made on examples of two of the most cited papers, {\it i.e.} by Brill and Goodman~\cite{BrG} and by Jackson~\cite{JDJ}.

Brill and Goodman (Sec. III of their work) prove the equivalence of the gauges by demonstrating the equality of the integrals representing ${\bf E}_C$ and ${\bf E}_L$. Except for an erroneous assumption that the temporal and spatial variables are separated in the charge and current densities~\footnote{This cannot be true for the classical charges. The charge density for these particles in arbitrary motion is given by $\rho({\bf r},t) = \delta[{\bf r}-{\bf r}_0(t)]$, where ${\bf r}_0$ is the coordinate of the particle trajectory. For such a dependence, $\rho({\bf r},t)=\rho({\bf r},\omega)e^{i\omega t}$ cannot be used.} the authors ignore the difference in the integration domains. 

Let us consider the following system: a classical charge is at rest at point $O$ for $t< 0$. The charge creates the Coulomb potential that is established in all space, which provides the equivalence of the electric field in both gauges. At the instant $t=0$ the charge begins to move along the $x$ axis in such a way that it approaches the point $A$ at $t^*$ (Fig. 1). During this process, the conditions of gauge equivalence can change, because the potentials created by the charge in its motion also change. Let us state a question: how to determine these potentials. 

\begin{figure}[htbp]
\begin{center}
\includegraphics[bb = 0 0 548 337, scale=0.6]{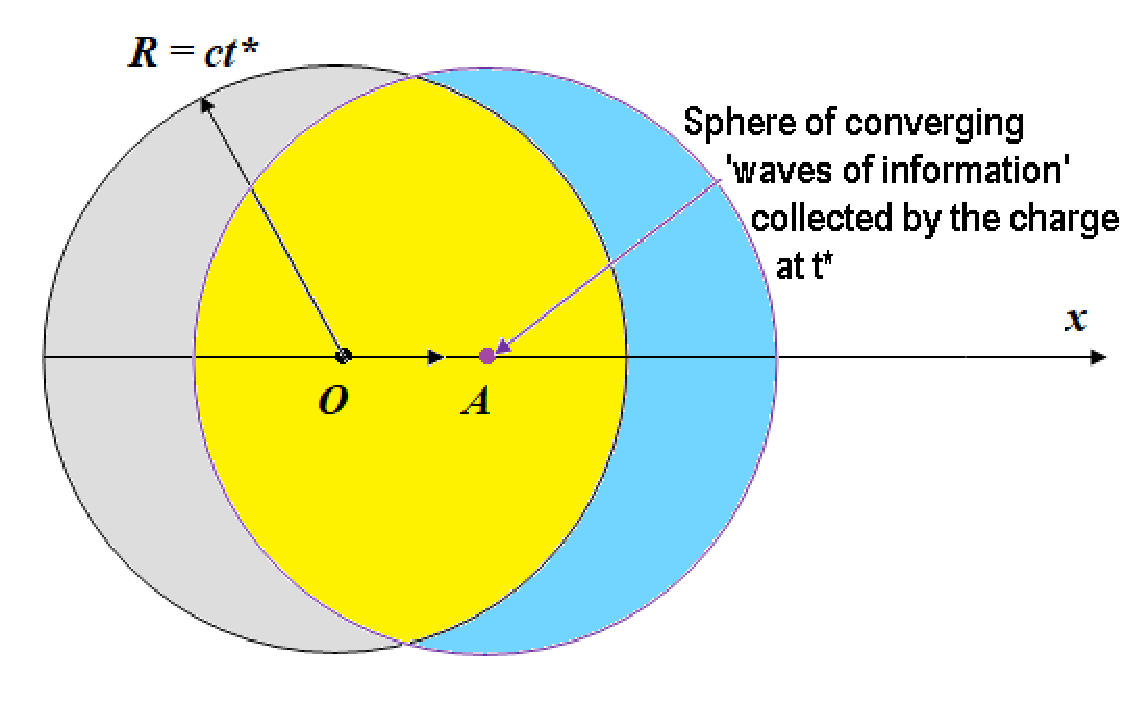}
\caption{The system considered to demonstrate what integration domains are created by a single charge in two gauges. $OA$ is  the path of the charge.}
\end{center}
\end{figure}

Since for an arbitrary law of motion of the charge $\varphi_L$ and ${\bf A}_\Phi$ are represented by the retarded integrals, it is necessary to determine the integration domains for them. These domains must be spherical regions with the center at $A$. The explanation why the integration domain has such a shape is given in ch. 19-1 of~\cite{PPh} -- 'wave of information' about the values of the potentials starts at $t=0$ from the surface of the sphere with radius $R=ct^*$ and converges with the speed of light to p. $A$, where all information is collected at the time $t^*$.

For the integral of ${\bf A}_\Phi$, the integration domain is the sphere with radius $R=ct^*$ and centered at p. $A$, since the non-local source of the integral -- the derivatives of the Coulomb scalar potential -- exists in all space. But this is not the case for the integral of $\varphi_L$. If at $t<0$ $\varphi_L$ is the Coulomb potential established in all space, when the charge moves, a new scalar potential is established in the surrounding space as a spherical wave expanding at the speed of light. At time $t^*$, this scalar potential created by the moving charge will be established in the spherical region of radius $R=ct^*$ centered at p. $O$ (the light purple region in Fig. 1). For the retarded integral of $\varphi_L$, the region of integration must be the same sphere centered at p. $A$. However, the expanding waves of the establishing $\varphi_L$ are not able to reach the region of space marked by the blue color in Fig. 1. Therefore, the integration domain for this potential is the region marked by the yellow color. 

One can see that these features of the integration domains of the retarded integrals are ignored in the paper of Brill and Goodman. Therefore their final statement that '{\it reinserting the time dependence $e^{-i\omega t}$, one then sees that} ${\bf E}_C={\bf E}_L$' cannot be accepted as correct.  

It should be noted that similar incorrect assumption is given in~\cite{JDJ}, {\it i.e.} that the temporal and spatial variables are separated in the charge density $\rho$ (Eq.~(3.11)). Correct rule of computation of $\partial \rho/\partial t$ for the the charge density of the form $\rho({\bf r},t) = \delta[{\bf r}-{\bf r}_0(t)]$ is $\left(\nabla\rho \cdot \dfrac{\dd r_0(t)}{\dd t}\right)$. But it should essentially complicate the calculations. 

But a more serious error is the ignorance of the difference of integration domains. In~\cite{JDJ} the author chooses another criterion of gauge equivalence, {\it i.e.} the difference between the so-called {\it gauge functions} $\chi_C$ and $\Psi$.  Partial derivatives of these gauge functions give the differences $\varphi_C-\varphi_L$ and ${\bf A}_\Phi={\bf A}_C-{\bf A}_L$. So the equivalence of the electric fields in two gauges should follow from the relation $\Psi({\bf r},t)=\chi_C({\bf r},t)+const$ (Eq.~(4.5) of~\cite{JDJ}).  

Let us note, however, that both gauge functions are represented by the retarded integrals with corresponding integration domains. Since the integration domain for $\chi_C'$, which is determined by $\varphi_L$, coincides with the integration domain of this potential itself, and the integration domain of $\Psi$ coincides with the integration domain of ${\bf A}_\Phi$, these domains are different, as follows from the above consideration. But the author of~\cite{JDJ} presents only integrals expressed in a general form, without any analysis at what limits the evaluation of these integrals should be made. 

Since in all works on this subject the retarded integrals for $\varphi_L$ and ${\bf A}_\Phi$ should be computed, the authors of these works had to consider the difference in the integration domains for these quantities. Unfortunately, no one of them made the corresponding analysis of the integration domains. Based on this fact alone, it can be concluded that the proofs of the gauge equivalence presented in these works are at least incomplete.

The only exception is work of Hnizdo~\cite{Hn}. But this author considers the system where the charge moves uniformly from $x=-\infty$ to $x=+\infty$. In this case, due to the translational symmetry of the system, the problem of different integration domains does not exist. However, if the condition $\dd {\bf v}/\dd t$ were broken, the problem of the difference in the integration domains would be real. 

\section{Conclusions}

In this work, methods for solving Maxwell's equations are analyzed. The first method is based on the direct derivation of the wave equation for the EM fields without the introduction of auxiliary quantities, the potentials. Although this method eventually gives the correct result~\footnote{Applying the operators $\partial_t$ to the first of the equations.~(\ref{WaEq}) and $\nabla$ to the second of these equations, and then term-by-term summation gives the wave equation for the electric field. But this equation is a consequence of two wave equations for the potentials.}, its derivation is not correct from a mathematical point of view (Sec. 2). 

The most used method for solving the Maxwell equations is based on reducing them to two wave equations for the potentials, solving them, and calculating the EM fields from the resulting potentials. In this method, deriving the wave equations is a rigorous mathematical procedure. But separating the equations for $\varphi$ and ${\bf A}$ requires using a coupling condition for these potentials, or a gauge condition. The mere presence of this gauge condition yields an infinite number of solutions. Then the only way to achieve uniqueness of a solution to the Maxwell equations is to show that calculating the partial derivatives of the resulting potentials using the rule~(\ref{P}) yields a unique expression for the electric field. Therefore, the problem of uniqueness of a solution to Maxwell's equations is reduced to the problem of gauge equivalence: to have the desired uniqueness, one must prove that the EM fields calculated in two different gauges are described by identical expressions.  

Unfortunately,  the evaluation of the expressions for potentials in two gauges faces obstacles that cannot be overcome. The scalar potential in the Lorentz gauge $\varphi_L$ is represented by a retarded integral, which cannot be calculated even for $\varphi_L$ created by a single charge -- in the case of arbitrary motion of the latter. The scalar potential in the Coulomb gauge, the second most frequently used, is the instantaneous Coulomb potential, which is easy to calculate. But the difference of vector potentials in the two gauges, ${\bf A}_\Phi$, is also represented by a retarded integral.  

Since the retarded integrals cannot be evaluated, they can be transformed in such a way that something like the relation~(\ref{Rel}) is established for them. Most authors of works on this topic adhere to this approach. But when transforming the quantities included in~(\ref{Rel}), all these authors ignore one property of these retarded integrals, {\it namely} the domains of integration of these integrals are different, which is shown in Sec. 4. Thus, when the authors transform the expressions, but omit the transformation of the domains of integration it is serious lack in proving the gauge equivalence.  

It may be objected that a system in which the difference in integration domains exists is very specific, and in other systems the integration domains may be the same. But considering the gauge equivalence one should introduce the 'initial conditions', {\it i.e.} such fields produced by sources that these fields are equal in each gauge. The fields of the charge, which has been at rest for a sufficiently long time $-\infty<t<0$ for this Coulomb potential to have been established in all space, satisfy these conditions. If at $t=0_-$ the sources are moving, one should determine what the initial EM is at that instant. In other words, the problem of determining the EM fields at the time $t>0$ is 'shifted' to the problem of determining the EM fields at the initial time.  

It should be said that the difference in the integration domains is not caused by specific properties of the system under consideration. It is caused by the difference in the sources of the retarded integrals. For $\varphi_L$ the source is a point-like object, so establishing the potential when the source changes its velocity occurs like spherical waves expanding from the source. For ${\bf A}_\Phi$ the source is initially extended in all space and no time is needed to establish the potential ${\bf A}_\Phi$.

Finally, it should be noted that the retarded integrals could be calculated in some specific cases (works~\cite{Hn, R-G}), but this is only possible for specific laws of charge motion. For arbitrary motion of the charge the calculation of the integrals is impossible. This does not mean that the EM fields calculated in two different gauges must be different. The analysis presented in this article allows us to conclude that it is impossible to state that the EM fields calculated in these gauges must be equal.

As a consequence, one cannot say that the Maxwell equations have a unique solution, because the gauge condition connecting the potentials can be chosen by an infinite number of cases. 
The uniqueness of the solutions of the Maxwell equations is provided if it is accepted that the only gauge that can be used in electrodynamic calculations is the Lorenz gauge. It is in agreement with statement by Cote and Johnson in~\cite{CJ} 'to abandon the gauge concept in classical 
electromagnetism'.

\section*{Appendix}

In his book~\cite{Jef} Jefimenko states that the solution of the equation
\begin{equation}
-\left[\nabla\times[\nabla\times{\bf E}]\right]-\frac{\partial ^2{\bf E}}{c^2\partial t^2}=\frac{4\pi\partial{\bf j}}{c^2\partial t}\,.\label{A1}
\end{equation}
is 
\[
{\bf E}({\bf r};t)=-\frac{1}{4\pi}\int \frac{\left[\nabla'(\nabla'\cdot){\bf E}+
(1/c)\partial {\bf J}({\bf r}';t')/\partial t' \right]_{ret}}
{|{\bf r}-{\bf r}'|}\dd^3 r' \,,
\]
where the quantities in the square brackets depend on the retarded time $t'$.

The author does not present a proof of this statement. Instead, he refers to Richardson's '{\it The electron theory of matter}' and McQuisten's '{\it Scalar and vector fields}', Ch. 12.3 'The retarded potentials'. 

Richardson presents the derivation of the wave equation for the electric field from the Maxwell equations (p. 187 of~\cite{Rich}). Let us shortly analyze this derivation.

Differentiation of Eq.~(\ref{Mxw2}) with respect to $t$ gives
\begin{equation}
\frac{\partial^2{\bf E}}{\partial t^2} + \frac{\partial}{\partial t}\left({\bf v}\rho \right)=
c\frac{\partial}{\partial t}[\nabla\times{\bf H}] \,,\label{A2}
\end{equation}
where, as the author assumes, ${\bf j}={\bf v}\rho$.

Then one more Maxwell equation is applied to eliminate the magnetic field from~(\ref{A2}) which gives
\begin{equation}
\frac{\partial^2{\bf E}}{\partial t^2} + \frac{\partial}{\partial t}\left({\bf v}\rho \right)=
-c^2\left[\nabla\times[\nabla\times{\bf E}]\right] \,.\label{A3}
\end{equation}
Using the vector identity,
\[
\left[\nabla\times[\nabla\times{\bf E}]\right]=\nabla\left( \nabla\cdot{\bf E}\right)
-\nabla^2{\bf E}\,,
\]
one can transform Eq.~(\ref{A3}) to
\begin{equation}
\frac{1}{c^2}\frac{\partial^2{\bf E}}{\partial t^2} + \frac{1}{c^2}\frac{\partial}{\partial t}\left({\bf v}\rho \right)=
\nabla^2{\bf E}- \nabla\left( \nabla\cdot{\bf E}\right) \,.\label{A4}
\end{equation}

consideers 
But Richardson considers the radial wave equation of Kirchhoff but not the equation of form (\ref{A1}). The only equation to Eq.~(\ref{A1})  in McQuisten's book is Eq.~(12.40)~\cite{McQ} but this equation is written for the vector potential in the Coulomb gauge and cannot be applicable to the electric field.

In fact, correctness of Eq.~(\ref{A1}) means that if one adds $\nabla'(\nabla'\cdot{\bf E})$
 to the left and right hand sides of Eq.~(\ref{A1}), the  {\it lhs} of this equation acquires a form of the (true) wave equation. Then $\nabla'(\nabla'\cdot{\bf E})$ of the {\it rhs} must be treated as a part of the source. But according to the Duhamel principle, terms in the left and right hand sides of any inhomogeneous PDE have different meaning; the terms in the {\it rhs} is like a set of initial conditions for homogeneous PDE.
 
Since Jefimenko cites the book of Richardson, it would be reasonable to note that this author presents a derivation

\end{document}